\documentstyle[12pt]{article}

\newcommand{\be}{\begin{equation}}
\newcommand{\ee}{\end{equation}}
\newcommand{\ba}{\begin{eqnarray}}
\newcommand{\ea}{\end{eqnarray}}
\begin{document}
%%%%%%%%%%%%%%%%%%%%%%%%%%%%%%%%%%%%
%%%%%%%%%%%%%%%%%%%%%%%% FRONT PAGE %%%%%%%%%%%%%%%%%%%%%%%%%%%%%%%%%%%%%
\begin{titlepage}
\hbox{\hskip 12cm CPHT-S.668.1098  \hfil}
\hbox{\hskip 12cm hep-th/9810214 \hfil}
\hbox{\hskip 12cm \today }
\vskip .8cm
\begin{center}  {\Large  \bf  Non-tachyonic open descendants 
\\
\vskip 24pt of the 0B string theory}
 
\vspace{1.2cm}
 
{\large \large Carlo Angelantonj\footnote{e-mail: 
angelant@cpht.polytechnique.fr}}

\vspace{.3cm}

{\it Centre de Physique Th\'eorique}

{\it Ecole Polytechnique}

{\it F-91128 Palaiseau, France}

\end{center}
\vskip .6cm

\abstract{We use the {\it crosscap constraint} to construct open
descendants of the 0B string compactified on $T^6 /Z_3$ and on $T^4
/Z_2$ free of tachyons both in the closed and in the open
unoriented sectors. In four dimensions the construction results in a
Chan-Paton gauge group ${\rm U} (8) \otimes {\rm U} (12) \otimes {\rm
U} (12)$ with three generations of chiral fermions in the
representations $(\overline{8} , 1 ,\overline{12}) \oplus (8, 12,1)
\oplus (1 ,\overline{66} ,1) \oplus (1,1,66)$.}
 
\vfill \end{titlepage}
%%%%%%%%%%%%%%%%%%%%%%%%%%%%%%%%%%%%%%%%%%%%%%%%%%%%
\makeatletter
\@addtoreset{equation}{section}
\makeatother
\renewcommand{\theequation}{\thesection.\arabic{equation}}
\addtolength{\baselineskip}{0.3\baselineskip} 
%%%%%%%%%%%%%%%%%%%%%%%%%%%%%%%%%%%%%%%%%%%%%%%%%%%%

\section{Introduction}

It has long been known  that,
besides the five supersymmetric strings,
there are a number of non-supersymmetric theories in ten dimensions
\cite{nonsusy}. 
These can be 
obtained as $Z_2$ orbifolds of the supersymmetric ones, where the
$Z_2$ generator $(-)^F$ is related to the total fermion number $F$,
and is accompanied by an
action on the gauge group in the case of the heterotic strings. 
Particularly interesting for what concerns our purposes are the
two tachyonic 0A and 0B theories, that descend from the type IIA and
IIB superstrings. Using the characters of affine
${\rm SO} (8)$ at level one, their partition functions read
\ba
{\cal T}_{{\rm 0A}} &=& |O_8 |^2 + |V_8|^2 + S_8 \bar C_8 + C_8 \bar
S_8 \,,
\label{fp0a} 
\\
{\cal T}_{{\rm 0B}} &=& |O_8|^2 + |V_8|^2 + |S_8 |^2 + |C_8|^2 \,.
\label{fp0b}
\ea
The NS-NS sector, common to both theories, includes a tachyon,
together with
a metric tensor, a dilaton and a rank 2 antisymmetric tensor at
the massless level. The R-R sectors are different. At the massless
level the 0A string contains two vectors and two 3-forms, thus simply
doubling the type IIA R-R massless states, while the 0B theory 
contains two scalars, two 2-forms and one 4-form without definite
chirality. Being non chiral, both theories are anomaly free in ten
dimensions. 

One common and important feature of the 0A and 0B string theories is
that they are symmetric under the interchange of left and right
modes. 
We can then apply the procedure developed in \cite{cargese,bsl} to mod out
both theories by the world-sheet parity $\Omega$ and construct 
open descendants. 
In ten dimensions this has been done long ago by Bianchi and Sagnotti 
\cite{bsl} (see also \cite{bergman}). If we choose a ``standard'' action
for the $\Omega$ projection that symmetrises the NS-NS sector
and antisymmetrises the R-R one, typically we get open descendants
tachyonic both in the closed and open unoriented sectors. 
But this is not the only choice available for the action of the
world-sheet parity. In \cite{fps} it has been shown that 
one can suitably modify the Klein bottle projection in a way 
consistent with what has been called the {\it crosscap constraint}.
Typically, for supersymmetric theories these ``exotic'' $\Omega$
projections result in open descendants without the open
unoriented sector \cite{gep}, since the Klein bottle amplitude
in the transverse channel does not contain any massless states,
and thus does not contribute to tadpoles. As a result, one is forced to
set to zero the Chan-Paton multiplicities in order to get a consistent
model. Fortunately, this is not the case for the non-supersymmetric
theories. In the 0B string one can use the sign ambiguities 
in the Klein bottle projection to remove the tachyons from the
spectrum. In ten dimensions this has been shown by Sagnotti
\cite{susy95} (see also \cite{ik5}). 
The final result is a chiral closed unoriented massless
spectrum consisting of a metric tensor, a dilaton, a rank 2
antisymmetric tensor, a scalar and a (anti)self-dual 4-form, and 
in the open unoriented sector
a ${\rm U} (32)$ gauge group with Majorana-Weyl fermions in the $496 \oplus 
\overline{496}$ representations . In this
case the anomaly cancellation involves the R-R (anti)self-dual
4-form and the 8-form dual to the R-R scalar, in addition to the R-R
2-form, in a generalized Green-Schwarz mechanism \cite{susy95}.
In the 0A string theory it is not possible to modify the Klein bottle
projection, and thus one can not
remove the tachyons from the spectrum.

At this point one may ask if this result can be generalized 
lower dimensions. This is precisely the aim of this letter. We
will study orbifold compactifications of the 0B string theory to six
and four dimensions and will look for ``exotic'' Klein bottle
projections in order to remove tachyons whenever this is possible. 
In section 2 we discuss 
compactifications of the 0B string to four dimensions and
construct its open descendants, both
tachyonic and non-tachyonic, in the $T^6 /Z_3$ case. 
Section 3 is devoted to the
six-dimensional  $T^4 /Z_2$ case.

\section{Four dimensional compactifications}

Let us start by considering four-dimensional compactifications on
orbifolds $T^6 / Z_N$ ($N=3,4,6,7,8,12$) \cite{orb}. For a generic $N$,
the 0B string partition function associated to the geometrical action
of the orbifold generators may be written as
\ba
{\cal T} &=& {1 \over N} \sum_{\alpha ,\beta = 0}^{N-1} \,
n_{\alpha,\beta} \biggl\{ 
\gamma_{\alpha ,\beta} \, \bar\gamma _{-\alpha,-\beta} + 
\delta_{\alpha ,\beta} \, \bar\delta _{-\alpha,-\beta} +
\eta_{\alpha ,\beta} \, \bar\eta _{-\alpha,-\beta} +
\epsilon_{\alpha ,\beta} \, \bar\epsilon _{-\alpha,-\beta} \biggr\}
\times
\nonumber 
\\ 
& & \qquad \qquad \times \prod_{i=1}^{3}
(\phi^{i}_{\alpha,\beta} \, \bar\phi^{i}_{-\alpha
,-\beta}) 
\left[ \Lambda^{i}_{2} (\tau , \bar\tau ) \right]_{(\alpha,\beta)=(0,0)} \,.
\label{4do0b}
\ea
For a given twist $(t_1 , t_2 , t_3)$ on the internal torus $T^6
= T^2 \times T^2 \times T^2$, the contribution of the world-sheet
fermions is
\ba
\gamma_{\alpha ,\beta} &=& {1\over 2\eta^3} \left\{ (O_2 + V_2 )
\prod_{i=1}^{3}\, \theta\left[{}^{\alpha t_i}_{\beta t_i}\right] +
(O_2 - V_2) \prod_{i=1}^{3} \, \theta \left[ {}^{\ \ \alpha t_i}_{1/2
+ \beta t_i}\right] \right\} \,,
\nonumber
\\
\delta_{\alpha ,\beta} &=& {1\over 2\eta^3} \left\{ (O_2 + V_2 )
\prod_{i=1}^{3}\, \theta\left[{}^{\alpha t_i}_{\beta t_i}\right] -
(O_2 - V_2) \prod_{i=1}^{3} \, \theta \left[ {}^{\ \ \alpha t_i}_{1/2
+ \beta t_i}\right] \right\} \,,
\nonumber
\\
\eta_{\alpha ,\beta} &=& {1\over 2\eta^3} \left\{ (S_2 + C_2 )
\prod_{i=1}^{3}\, \theta\left[{}^{1/2+\alpha t_i}_{\ \ \beta t_i}\right] +
i (S_2 - C_2) \prod_{i=1}^{3} \, \theta \left[ {}^{1/2+\alpha t_i}_{1/2
+ \beta t_i}\right] \right\} \,,
\nonumber
\\
\epsilon_{\alpha ,\beta} &=& {1\over 2\eta^3} \left\{ (S_2 + C_2 )
\prod_{i=1}^{3}\, \theta\left[{}^{1/2+\alpha t_i}_{\ \ \beta t_i}\right] -i
(S_2 - C_2) \prod_{i=1}^{3} \, \theta \left[ {}^{1/2+\alpha t_i}_{1/2
+ \beta t_i}\right] \right\} \,,
\nonumber
\ea
while the contribution of the bosonic coordinates is 
\be
\phi^{i}_{\alpha , \beta} = 2 \sin (\beta t_i \pi) 
{\eta \over \theta \left[ {}^{1/2 + \alpha t_i}_{1/2 + \beta t_i}
\right]}\quad {\rm and} \quad \phi^{i}_{\alpha,\beta} = {1 \over \eta^2}
\ \ {\rm if}\ \ \alpha t_i , \beta t_i \in Z\,,
\ee
with $\eta$ the Dedekind function. 
The notation $[\Lambda^{i}_{2}]_{(\alpha
,\beta)=(0,0)}$ means that the lattice sums are present only when the
orbifold generator acts trivially on a particular
direction. Finally, the numbers $n_{\alpha ,\beta}$ represent the
fixed point multiplicities as given by the Lefschetz theorem. 

From the partition function (\ref{4do0b}) one can then extract the
spectrum of the theory. For generic $N$, the twisted sector includes 
a tachyon, directly related to the ten dimensional tachyon of
the 0B theory, that is invariant with respect to the action of the
orbifold. For $N\not= 3,4$ one also finds tachyons in the twisted
sectors. Their masses, however, are smaller in absolute value than that of the
untwisted tachyon. The orbifolds $Z_{3,4}$ are an exception, since
rather than twisted tachyons the spectrum contains
additional massive (for the $Z_3$ case) or massless (for the $Z_4$ case)
scalars. These states are Kaluza-Klein excitations of the untwisted
tachyon. The contribution of the internal compact scalar Laplacian to the 
four-dimensional field equations simply cancels the negative tachyonic
``mass'', leading to massless or massive states. For these twisted
states, the eigenvalues of the
internal Laplacian are independent of any
moduli and thus the mechanism takes place at every point in 
moduli space.

Since the theory is still left-right symmetric, one can
mod out by $\Omega$ and derive open descendants
\cite{ps,gp}. Typically, for the ``standard'' Klein bottle projection the
final result is a Chan-Paton gauge group that somehow doubles the one
associated with the same orientifold of the type IIB superstring, 
while the low-lying excitations inevitably contain tachyons
both in the closed and in the open unoriented sectors. 
What happens if we use the sign ambiguities left over from the 
crosscap constraint to change the Klein bottle projection? 
For $Z_N$ orbifolds with $N\not= 3,4$ nothing
interesting happens. The orthogonal and/or symplectic gauge groups
become unitary, the untwisted open and closed tachyons are thrown 
away, but the twisted closed ones still survive. 
This is due to the fact that for a geometrical action of
the orbifold generators, the associated partition function belongs to
the family of charge conjugation modular invariants. 
This means that states in the $\theta$-twisted sector are
paired with states in the $\theta^{-1}$-twisted one. As a result, they
do not contribute to the Klein bottle projection for any choice of
signs unless $\theta = \theta^{-1}$. The net number of tachyons
is then simply halved in the open descendant and there is no way to
eliminate them. The case is completely different for the $Z_{3,4}$
orbifolds. As we have said in the previous paragraphs, they contain only
the untwisted tachyon, and therefore one can hope to eliminate it using the
``exotic'' Klein bottle projection. 
For this reason we will now consider the parameter space orbifold of
the 0B theory on the $T^6 /Z_3$ orbifold in some detail.

The 0B string partition function is simply given by (\ref{4do0b}) with 
$N=3$ and $n_{0,\beta} = 1$, $n_{1,\beta} = n_{2,\beta} =27$. The 
low-lying excitations comprise a tachyon, a metric tensor, two
abelian vectors and 60 scalars from the untwisted sector, while
each of the 27 fixed points in the twisted sectors contribute 6
additional scalars, for a total of 162 scalar fields. 

In constructing the open descendants, one starts by halving the torus 
amplitude. Since the $Z_3$ action of the target space twist is L-R
symmetric, the ``standard'' Klein bottle amplitude is
\be
{\cal K} = {1 \over 6} \sum_{\beta=0,1,2} \left( \gamma_{0,\beta} +
\delta_{0,\beta} - \eta_{0,\beta} - \epsilon_{0,\beta} \right) \,
\phi_{0,\beta} \, \left[ P (2 i \tau_2 ) \right]_{\beta =0}\,,
\label{stkb4}
\ee
where $\phi_{0,\beta} = \phi^{1}_{0,\beta} \phi^{2}_{0,\beta} 
\phi^{3}_{0,\beta}$, and $[P]_{\beta =0}$ means that the lattice sum
over the KK momenta
is present only for $\beta =0$. 
Eq. (\ref{stkb4}) contains only  the conventional sum over the momentum
lattice since, for generic values of the moduli of the internal torus, 
the condition ${p_L} =\omega p_R$ ($\omega =e^{2 i \pi/3}$) 
does not have any non-trivial
solutions \cite{chiral}. 
One may thus anticipate that the open sector includes only 
Neumann charges associated with the D9-branes.  
This should be contrasted with the $Z_2$ case that we will consider in
the next section, where additional contributions to ${\cal K}$ signal the
appearance of Dirichlet charges in the open spectrum \cite{ps,gp}. 
The light states in the projected closed-string
spectrum thus comprise the untwisted tachyon, the metric tensor and
30 massless scalars from the untwisted sector, as well as 
81 massless scalars from the twisted sectors.

The twisted sector of the world-sheet orbifold, to be identified with the
open-string spectrum, starts with the annulus amplitude
\ba
{\cal A} &=& {1 \over 6}  \sum_{\beta =0,1,2}  \left\{ 2
(M_{\chi}^{(\beta)} M_{\xi}^{(\beta)} + M_{\sigma}^{(\beta)}
M_{\tau}^{(\beta)} )\, \gamma_{0,\beta} \phi_{0,\beta} \,+ \right.
\nonumber
\\
& & \qquad+\left[ (M_{\chi}^{(\beta)})^2 +
(M_{\xi}^{(\beta)})^2 +(M_{\sigma}^{(\beta)})^2 +(M_{\tau}^{(\beta)})^2
\right] \,  \delta_{0,\beta} \phi_{0,\beta}\, +
\nonumber 
\\
& & \qquad - 2 (M_{\chi}^{(\beta)} M_{\tau}^{(\beta)} +
M_{\xi}^{(\beta)} M_{\sigma}^{(\beta)} ) 
\,  \eta_{0,\beta} \phi_{0,\beta}+
\nonumber
\\
& & \qquad \left.
-  2 (M_{\chi}^{(\beta)} M_{\sigma}^{(\beta)} +
M_{\xi}^{(\beta)} M_{\tau}^{(\beta)} ) 
\,  \epsilon_{0,\beta} \phi_{0,\beta}\, 
\right\} \, \left[ P(i\tau_{{\rm A}}/2) \right]_{\beta =0} \,,
\label{sa4}
\ea
whose $\Omega$ projection is completed by the M\"obius strip amplitude
\be
{\cal M} = - {1\over 6} \sum_{\beta=0,1,2} (M_{\chi}^{(-\beta)} +
M_{\xi}^{(-\beta)} + M_{\sigma}^{(-\beta)} +
M_{\tau}^{(-\beta)} )\, \hat\delta_{0,\alpha} \hat\phi_{0,\alpha} 
\left[ \hat P (i \tau_{{\rm M}} /2 + 1/2) \right]_{\beta =0}
\label{sms4} \,,
\ee
where we have defined a suitable set of real ``hatted'' characters in the
M\"obius amplitude \cite{bsl}. The coefficients $M_{i}^{(\beta )}$ are
\be
M_{i}^{(\beta )} = n_i + \omega ^\beta m_i + \bar\omega^\beta 
\bar m _i \,, \qquad (i=\chi , \xi, \sigma ,\tau )\,,
\nonumber
\ee
where $n_i\,,\ m_i\,,\ \bar m_i$ are Chan Paton (CP) charges.

Tadpole cancellation in the transverse channel leads to the two conditions
\ba   
M^{(0)}_{\chi} + M^{(0)}_{\xi} + M^{(0)}_{\sigma} + M_{\tau}^{(0)} &=&
64 \,,
\nonumber \\  
M^{(i)}_{\chi} + M^{(i)}_{\xi} + M^{(i)}_{\sigma} + M^{(i)}_{\tau} &=&
-8 \,,  \qquad (i=1,2)
\label{stad4}
\ea   
related to untwisted and twisted massless exchanges respectively.
All other tadpoles are trivially satisfied as a result of the numerical
identification of conjugate charges.
Together with (\ref{sms4}) and (\ref{sa4}), eqs. (\ref{stad4})  
yield a semi-simple CP gauge group
\be
G_{{\rm CP}} = \left[ {\rm SO} (n) \otimes {\rm SO} (8-n) \otimes {\rm
U} (m) \otimes {\rm U} (12-m) \right]^{\otimes 2} \,,
\label{scp4}
\ee
containing 8 factors, as well as
\ba
1\ {\rm tachyon} &\in& (\bar F _3 , F _7 ) \oplus
( F_4 , \bar F_8) \oplus (F _3,\bar F_7 ) 
\oplus (\bar F_4 , F_8 ) \oplus (F_1 ,F_5 ) 
\oplus (F_2,F_6 ) \,,
\nonumber
\\
3\ {\rm scalars} &\in& \bar A _3 \oplus
\bar A_7  \oplus \bar A_4 
\oplus \bar A _8 \oplus (F_1 , F_3 ) \oplus
(F_5 , F_7)
\oplus (F_2 ,F_4 ) \oplus (F_6 ,F_8)
\oplus ({\rm c.c.}) \,,
\nonumber
\\
1\ {\rm L\ fermion} &\in& (F_7,\bar F_8) \oplus (\bar F _3
, F_4 ) \oplus (\bar F_7 , F_8) \oplus (F_3 ,\bar F _4
) \oplus (F_1,F_2) \oplus (F_5,F_6) \oplus
\nonumber
\\
& & \oplus (\bar F_4 ,F_7) \oplus (F_4 ,\bar F_7) \oplus
(\bar F_3 ,F_8) \oplus (F_3,\bar F_8) \oplus
(F_2,F_5 ) \oplus (F_1,F_6) \,,
\nonumber
\\
3\ {\rm L\ fermions} &\in& (\bar F_7 ,\bar F_8) \oplus
(\bar F_3 ,\bar F_4 ) \oplus (F_5,F_8) 
\oplus (F_2,F_3) \oplus (F_6,F_7) \oplus (F_1,F_4) \oplus
\nonumber
\\
& & \oplus (F_4,F_7) \oplus (F_3,F_8) \oplus (\bar F_4 ,F_5) 
\oplus (F_2,\bar F_7) \oplus (\bar F _3, F_6 ) \oplus (F_1,\bar F_8) \,,
\nonumber
\ea
where $F_i\ (A_i )$ stands for the fundamental (antisymmetric)
representation of the $i$-th factor group
and the bars refer to the conjugate representations. The model is free of
non-abelian anomalies as a consequence of the cancellation of twisted
tadpoles.

Now we can use the results of \cite{fps} to modify the Klein
bottle projection and to eliminate the tachyons. There are two
choices that one can make, but they differ by an overall change of
chirality for the fermions in the open unoriented sector. We
can thus concentrate on one of the two possibilities, say
\be
{\cal K}' = {1 \over 6} \sum_{\beta =0,1,2} (-\gamma_{0,\beta} +
\delta_{0,\beta} -\eta_{0,\beta} + \epsilon_{0,\beta} ) \, \phi_{0,\beta}
\, \left[ P (2i\tau_2 )\right]_{\beta =0} \,.
\label{ekb4}
\ee
The massless states in the projected closed-string
spectrum comprise the metric tensor, 
30 scalars and an abelian vector from the untwisted sector, 
as well as 81 massless scalars from the twisted sectors.

The open unoriented one-loop amplitudes corresponding to
the projection (\ref{ekb4}) are given by
\ba
{\cal A} ' &=& {1 \over 6} 
\sum_{\beta =0,1,2} \left\{ 2\, (M^{(\beta)}_{1} \bar M^{(\beta)}_{2} +
\bar M^{(\beta)}_{1} M^{(\beta)}_{2} )\, \gamma_{0,\beta}
\phi_{0,\beta} + \right.
\nonumber
\\
& &\qquad + 2\, (M_{1}^{(\beta)} \bar M_{1}^{(\beta)} + 
M_2 ^{(\beta)} \bar M^{(\beta)}_{2} )\, \delta_{0,\beta}
\phi_{0,\beta} +
\nonumber
\\
& & \qquad - \left[ (M_{1}^{(\beta)})^2 + (\bar
M_{1}^{(\beta)})^2
+ (M^{(\beta)}_{2})^2 + (\bar M^{(\beta)}_{2})^2 \right]\,
\eta_{0,\beta} \phi_{0,\beta} +
\nonumber
\\
& & \qquad \left. - 2\, (M_{1}^{(\beta)} M_{2}^{(\beta)} +
\bar M_{1}^{(\beta)} \bar M_{2}^{(\beta)} )\, \epsilon_{0,\beta}
\phi_{0,\beta} \right\} 
\left[ P (i \tau_{{\rm A}} /2)\right]_{\beta =0} \,,
\label{ea4}
\ea
and
\be
{\cal M}'= {1\over 6}\sum_{\beta=0,1,2} (M_{1}^{(-\beta)} + \bar
M_{1}^{(-\beta)}
+ M_{2}^{(-\beta)} + \bar M_{2}^{(-\beta)} ) \, \hat\eta_{0,\beta} 
\hat\phi_{0,\beta}\, \left[\hat P (i\tau_{{\rm M}} /2 +
1/2)\right]_{\beta =0} \,,
\label{ems4}
\ee

In this second case, the coefficients that appear in the annulus and
M\"obius strip amplitudes are given by
\be
M_{k}^{(\beta)} = n_k + \omega^\beta m_k + \bar\omega^\beta p_k \,,
\qquad (k=1,2) \,,
\nonumber
\ee
and the CP charges are all complex, so that the gauge group is of the
form ${\rm U} (n) \otimes {\rm U} (m) \otimes {\rm U} (p)$. 
In order to extract
the dimensions of each factor and to check the consistency of the
model, one has to cancel tadpoles in the transverse channel. 
In this case, one does not have the option to set to zero the dilaton
tadpole, and the overall size of the gauge group is therefore not
determined. 
It has been shown in \cite{bachas} that this situation may lead to 
interesting progress in open string model building. 
However, there is unique choice of gauge group leading
to a tachyonic-free open spectrum. Therefore, the only
relevant tadpole conditions are given by
\ba
M_{1}^{(0)} + \bar M_{1}^{(0)} - M_{2}^{(0)} -\bar M ^{(0)}_{2} &=& 64
\,,
\nonumber
\\
M_{1}^{(i)} + \bar M_{1}^{(i)} - M_{2}^{(i)} -\bar M_{2}^{(i)} &=& -8
\,, \qquad (i=1,2) \,,
\label{etad4}
\\
M_{1}^{(i)} -\bar M_{1}^{(i)} - M_{2}^{(i)} +\bar M _{2}^{(i)} &=&
0\,,
\ \qquad (i=1,2)\,.
\nonumber
\ea
Together with (\ref{ea4}) and (\ref{ems4}), and putting to zero the
$n_2 \,,\, m_2 \,,\, p_2$ CP charges, one can see that the
solution
\be
G_{{\rm CP}} = {\rm U} (8) \otimes {\rm U} (12) \otimes {\rm U}
(12)\,,
\ee
of equations (\ref{etad4}) results in an open unoriented sector
free of tachyons and with additional charged massless matter given by
\ba
3\ {\rm scalars} &\in& (\overline{8} , 12 , 1) \oplus
(12,1,\overline{12}) \oplus
(1,\overline{12} , 12) \oplus ({\rm c.c.}) \,,
\nonumber
\\
1\ {\rm L\ fermion} &\in& (28,1,1) \oplus (\overline{28},1,1)
\oplus (1,12,12) \oplus (1,\overline{12} ,\overline{12}) \,,
\nonumber
\\
3\ {\rm L\ fermions} &\in& (\overline{8} ,1, \overline{12}) \oplus 
(8,12,1) \oplus
(1,\overline{66} ,1) \oplus (1,1,66) \,.
\nonumber
\ea
It is not difficult to see that the 
spectrum is anomaly free (aside from ${\rm U} (1)$ factors), as a 
result of the cancellation of twisted tadpoles.

In the construction of the previous models one could have also
considered compactifications in the presence of a quantized NS-NS
antisymmetric tensor background. It has been shown \cite{tensor}
that a $B_{\mu\nu}$ of rank $r$ reduces the dimension of the CP gauge
group by a factor of $2^{r/2}$. In our case three different choices
are allowed, $r=2,4,6$, i.e. one can introduce a non-vanishing tensor
background in one, two or three orbifold ``planes''. The resulting
massless open spectra, free of non-abelian anomalies, correspond to a
CP group ${\rm U} (8) \otimes {\rm U} (4) \otimes {\rm U} (4)$,
with three real scalars in the representations $(\overline{8} , 4,1)
\oplus (8,1,\overline{4} ) \oplus(1,\overline{4},4)\oplus ({\rm
c.c.})$, one left fermion in the representations $(36,1,1) \oplus
(\overline{36} ,1,1)\oplus (1,4,4) \oplus (1,\overline{4}
,\overline{4})$ and three left fermions in the representations
$(\overline{8},1,\overline{4}) \oplus (8,4,1)\oplus (1,\overline{10}
,1) \oplus (1,1,10)$ for the $r=2$ case, 
to a CP group ${\rm U} (4) \otimes {\rm U} (4)$,
with three real scalars in the representations $(\overline{4} , 4)
\oplus (4,\overline{4} )$, one left fermion in the representations 
$(4,4) \oplus (\overline{4} ,\overline{4})$ and three left 
fermions in the representations
$(\overline{6},1) \oplus (1,6)$ for the $r=4$ case, and, finally,
to a CP group ${\rm U} (4)$ with one left fermion in the
representations $10\oplus \overline{10}$ for the $r=6$ case.

\section{Six dimensional compactifications}

Six dimensional models display some similarities with the four dimensional
case. The orbifolds $T^4 /Z_{3,4,6}$ include in the light excitations
tachyonic states in the twisted sectors that can not be removed in the
open descendants by any choice of the Klein bottle projection. Only
for the $Z_2$ case the twisted sector is free from tachyons, and
we can then use the freedom left over from the crosscap constraint to find
a suitable $\Omega$ projection in the closed unoriented sector in
order to project them out. The partition function for the type 0B
string theory on the orbifold $T^4 /Z_2$ can be written 
\be
{\cal T} = {1 \over 2} \sum_{\alpha ,\beta =0,1} n_{\alpha ,\beta} \, \left[
|\gamma_{\alpha ,\beta} |^2 + |\delta_{\alpha ,\beta} |^2 + 
|\eta_{\alpha,\beta}|^2 + |\epsilon_{\alpha,\beta} |^2 \right]
\, |\phi_{\alpha ,\beta} |^2 \, \left[ \Lambda_4 (\tau ,\bar \tau)
\right] _{(\alpha ,\beta)=(0,0)} \,,
\label{6do0b}
\ee
where the amplitudes now read
\ba
\gamma_{\alpha ,\beta} &=& {1\over 2\eta^2} \left\{ (O_4 + V_4 )
\, \theta^2\left[{}^{\alpha /2}_{\beta /2}\right] +
(O_4 - V_4) \, \theta^2 \left[ {}^{\ \ \alpha /2}_{1/2
+ \beta /2}\right] \right\} \,,
\nonumber
\\
\delta_{\alpha ,\beta} &=& {1\over 2\eta^2} \left\{ (O_4 + V_4 )
\, \theta^2 \left[{}^{\alpha /2}_{\beta /2}\right] -
(O_4 - V_4) \, \theta^2 \left[ {}^{\ \ \alpha /2}_{1/2
+ \beta /2}\right] \right\} \,,
\nonumber
\\
\eta_{\alpha ,\beta} &=& {1\over 2\eta^2} \left\{ (S_4 + C_4 )
\, \theta^2 \left[{}^{1/2+\alpha /2}_{\ \ \beta /2}\right] -
(S_4 - C_4) \, \theta^2 \left[ {}^{1/2+\alpha /2}_{1/2
+ \beta /2}\right] \right\} \,,
\nonumber
\\
\epsilon_{\alpha ,\beta} &=& {1\over 2\eta^2} \left\{ (S_4 + C_4 )
\, \theta^2 \left[{}^{1/2+\alpha /2}_{\ \ \beta /2}\right] +
(S_4 - C_4) \, \theta^2 \left[ {}^{1/2+\alpha /2}_{1/2
+ \beta /2}\right] \right\} \,,
\nonumber
\ea
and
\be
\phi_{\alpha ,\beta} = \left( {2 \sin (\beta \pi /2) \eta \over \theta
\left[ {}^{1/2+\alpha /2}_{1/2+\beta /2} \right] } \right)^2 \,.
\nonumber
\ee
In this case the numerical coefficients are $n_{0,\beta}=1$ and
$n_{1,\beta} =16$, where 16 is the number of fixed points in the
orbifold $T^4 /Z_2$.
The low-lying excitations are then given by a tachyon, and, at
the massless level, by a metric tensor, 25 2-forms and 193 scalars.

In this case the open descendants contain more sectors, 
related to open strings with Dirichlet boundary conditions in the
internal directions, whose ends are stuck to the fixed points
of the orbifold. This translates into the inclusion of a new set of
charges \cite{ps,gp} associated to D5-branes. Limiting our
attention to the projection that removes the
tachyons both from the closed and the open unoriented sectors, the
relevant contributions are the Klein bottle amplitude
\be
{\cal K} =  {1 \over 4} \left( -\gamma_{0,0} + \delta_{0,0} -
\eta_{0,0} +\epsilon_{0,0} \right) \, \phi_{0,0} (P + W) + {16 \over 2}\,
\left( \gamma_{1,0} - \delta_{1,0} +\eta_{1,0} -\epsilon_{1,0} \right) \,
\phi_{1,0} \ ,
\label{ekb6}
\ee
the annulus amplitude
\ba
{\cal A} &=& {1\over 4} \sum_{\beta =0,1} \Biggl\{ \biggl[ 2
(N_{1}^{(\beta )} \bar N _{2}^{(\beta)} + \bar N _{1}^{(\beta )} N
_{2}^{(\beta)} ) \, \gamma_{0,\beta} + 2 (N_{1}^{(\beta )} \bar
N_{1}^{(\beta)} + N_{2}^{(\beta)} \bar N_{2}^{(\beta)} ) \,
\delta_{0,\beta} 
\nonumber 
\\& & \qquad \qquad - [(N_{1}^{(\beta)} )^2 + (\bar N_{1}^{(\beta)})^2 +
(N_{2}^{(\beta)})^2 + (\bar N _{2}^{(\beta)})^2 ] \, \eta_{0,\beta} 
\nonumber
\\
& & \qquad \qquad -
2 (N_{1}^{(\beta)} N_{2}^{(\beta)} +\bar N_{1}^{(\beta)} \bar
N_{2}^{(\beta)} ) \, \epsilon_{0,\beta} \biggr] \, \phi_{0,\beta} \,
[P]_{\beta =0}
\nonumber
\\
& & \qquad +\sum_{i,j=1}^{16} \biggl[ 2
(D_{1}^{(\beta )i} \bar D _{2}^{(\beta)j} + \bar D _{1}^{(\beta )i} D
_{2}^{(\beta)j} ) \, \gamma_{0,\beta} + 2 (D_{1}^{(\beta )i} \bar
D_{1}^{(\beta)j} + D_{2}^{(\beta)i} \bar D_{2}^{(\beta)j} ) \,
\delta_{0,\beta} 
\nonumber
\\
& & \qquad\qquad - (D_{1}^{(\beta)i} D_{1}^{(\beta)j} +  
\bar D_{1}^{(\beta)i} \bar D _{1}^{(\beta)j} +
D_{2}^{(\beta)i} D_{2}^{(\beta)j} + \bar D _{2}^{(\beta)i} 
\bar D _{2}^{(\beta)j} )  \, \eta_{0,\beta} 
\nonumber 
\\
& & \qquad \qquad -
2 (D_{1}^{(\beta)i} D_{2}^{(\beta)j} +\bar D_{1}^{(\beta)i} \bar
D_{2}^{(\beta)j} ) \, \epsilon_{0,\beta} \biggr] \, \phi_{0,\beta} \,
[W^{ij}]_{\beta =0}
\nonumber
\\
& & \qquad + 2 \sum_{i=1}^{16} \biggl[
(N_{1}^{(\beta)} \bar D _{2}^{(\beta)i} + \bar N _{2}^{(\beta)}
D_{1}^{(\beta)i} + \bar N_{1}^{(\beta)} D_{2}^{(\beta)i} +
N_{2}^{(\beta)} \bar D _{1}^{(\beta)i} ) \,\gamma_{1,\beta}
\nonumber
\\
& & \qquad \qquad 
+(N_{1}^{(\beta)} \bar D _{1}^{(\beta)i} + \bar N _{1}^{(\beta)}
D_{1}^{(\beta)i} + \bar N_{2}^{(\beta)} D_{2}^{(\beta)i} +
N_{2}^{(\beta)} \bar D _{2}^{(\beta)i} ) \,\delta_{1,\beta}
\nonumber
\\
& & \qquad\qquad
+(N_{1}^{(\beta)} D _{1}^{(\beta)i} + \bar N _{1}^{(\beta)}
\bar D_{1}^{(\beta)i} + N_{2}^{(\beta)} D_{2}^{(\beta)i} +
\bar N_{2}^{(\beta)} \bar D _{2}^{(\beta)i} ) \,\eta_{1,\beta}
\nonumber
\\
& & \qquad\qquad
+(N_{1}^{(\beta)} D _{2}^{(\beta)i} + N _{2}^{(\beta)}
D_{1}^{(\beta)i} + \bar N_{1}^{(\beta)} \bar D_{2}^{(\beta)i} +
\bar N_{2}^{(\beta)} \bar D _{1}^{(\beta)i} ) \,\gamma_{1,\beta}
\biggr] \, \phi_{1,\beta} \, \Biggr\} \ ,
\label{ea6}
\ea
and the M\"obius strip amplitude
\ba
{\cal M} &=& -{1\over 4} \Biggl\{ \biggl[ (N_{1}^{(0)} + \bar
N_{1}^{(0)} - N_{2}^{(0)} -\bar N _{2}^{(0)} ) \hat\eta_{0,0} \,
\hat\phi_{0,0} \, \hat P 
\nonumber
\\
& & \qquad - \sum_{i=1}^{16} ( D_{1}^{(0)i} + \bar D
_{1}^{(0)i} - D_{2}^{(0)i} - \bar D_{2}^{(0)i} ) \, \hat \eta_{0,0} \,
\hat \phi_{0,0} \, \hat W \biggr] 
\label{ems6} 
\\
& & + \biggl[ (N_{1}^{(0)} + \bar
N_{1}^{(0)} - N_{2}^{(0)} -\bar N _{2}^{(0)} ) 
- \sum_{i=1}^{16} ( D_{1}^{(0)i} + \bar D
_{1}^{(0)i} - D_{2}^{(0)i} - \bar D_{2}^{(0)i} ) \biggr] \hat
\eta_{0,1} \, \hat\phi_{0,1} \Biggr\} \,.
\nonumber
\ea
Here $W$ is the lattice sum restricted to winding states, and $W^{ij}$
has winding numbers shifted by the distance between the fixed points
$x_i$ and $x_j$, {\it e.g.} $ nR \to (n +x_i - x_j) R$. The notation
$[W^{ij}]_{\beta =0}$ means that this term is present for $\beta =0$,
while it is simply $\delta_{ij}$ if $\beta \not=0$. 
The coefficients in the amplitudes for the open unoriented
sector have the following expressions in terms of CP multiplicities
\be
N_{s}^{(\beta)} = n_s + (-)^\beta m_s \,, \qquad 
D^{(\beta)i}_{s} = p_{s}^{i} + (-)^\beta q_{s}^{i} \,, \qquad (\beta
=0,1\,,\ s=1,2\,,\ i=1,\ldots ,16)\,,
\ee
where $n,m$ ($p,q$) are associated to D9(5)-branes. In order to extract the
dimensions of the various factors of the CP gauge group, 
one has to cancel tadpoles in the
transverse channel. Again, the dilaton tadpole has to be
relaxed. The remaining tadpole conditions are
\ba
N_{1}^{(0)} + \bar N_{1}^{(0)} - N_{2}^{(0)} -\bar N _{2}^{(0)} &=&
64\,,
\nonumber
\\
\sum_{i=1}^{16} \left( D_{1}^{(0)i} + \bar D _{1}^{(0)i} -D_{2}^{(0)i}
- \bar D_{2}^{(0)i} \right) &=& -64 \,,
\label{etad6u}
\ea
from the untwisted sector, and
\be
{1 \over 4} \left( N_{1}^{(1)} - N_{2}^{(1)} + \bar N_{1}^{(1)} - \bar
N_{2}^{(1)} \right) + \left( D_{1}^{(1)i} - D _{2}^{(1)i} +\bar D_{1}^{(1)i}
- \bar D_{2}^{(1)i} \right) =0 \,, \qquad \forall i=1,\ldots ,16\,,
\label{etad6t}
\ee
from the twisted sector. Together with (\ref{ea6}) and (\ref{ems6}),
we can see that the solution
\be
G_{{\rm CP}} = \left[ {\rm U} (16) \otimes {\rm U} (16)
\right]^{\otimes 2} \,, 
\label{ecpgg6}
\ee
of equations (\ref{etad6u}) and (\ref{etad6t}) results in a model free
of tachyons both in the closed and in the open unoriented sectors. The
other massless excitations comprise a metric tensor, 4 anti-self-dual
2-forms, 20 self-dual 2-forms and 99 scalars from the closed sector,
as well as
\ba
4\ {\rm scalars} &\in& (16 , \overline{16} ; 1,1) \oplus
(\overline{16} , 16 ;1,1)
\oplus (1,1;16,\overline{16}) \oplus (1,1;\overline{16},16) \,,
\nonumber
\\
2\ {\rm L\ fermions} &\in& (136\oplus\overline{136},1;1,1) \oplus (1, 136
\oplus \overline{136} ;1,1) 
\nonumber
\\
& & \oplus (1,1;136 \oplus\overline{136},1)
\oplus (1,1;1,136\oplus\overline{136}) \,,
\nonumber
\\
2\ {\rm R\ fermions} &\in& (16,16;1,1) \oplus (\overline{16} ,
\overline{16} ; 1,1) \oplus (1,1;16,16) \oplus (1,1;\overline{16} ,
\overline{16}) \,,
\nonumber
\ea
from the open  untwisted sector, and
\ba
2\ {\rm scalars} &\in& (16,1;\overline{16},1) \oplus
(1,16;1,\overline{16}) \oplus (\overline{16} ,1;16,1) \oplus
(1,\overline{16} ;1,16) \,,
\nonumber
\\
1\ {\rm R\ fermion} &\in& (16,1;1,16) \oplus (1,16;16,1) \oplus
(\overline{16}, 1;1,\overline{16}) \oplus (1,\overline{16} ;
\overline{16} ,1) \,,
\nonumber
\ea
from the open twisted sector, where the fermions are symplectic Majorana-Weyl. 
In this case tadpole conditions guarantee the cancellation of the
irreducible part of the anomaly polynomial, 
while a generalized Green-Schwarz mechanism \cite{anom} is at work
to cancel the residual factorized anomaly:
\ba
{\cal I}_8 &=& -{1\over 8} \left[ {1\over 2} {\rm tr} R^2 - (
{\rm tr} F_{1}^{2} + {\rm tr} F_{2}^{2} + {\rm tr} F_{3}^{2}
+ {\rm tr} F_{4}^{2} ) \right]^2 +
\nonumber
\\
& & + {1\over 8} \biggl[ {1\over 4} ({\rm tr} F_{1}^{2} + 
{\rm tr} F_{2}^{2} - {\rm tr} F_{3}^{2} - {\rm tr} F_{4}^{2} )^2
+ {3\over 2} ({\rm tr} F_{1}^{2} + {\rm tr} F_{4}^{2} )^2
\nonumber
\\
& & \qquad + {3\over 2} ({\rm tr} F_{2}^{2} + {\rm tr} F_{3}^{2} )^2
+ {5\over 4} ({\rm tr} F_{1}^{2} -
{\rm tr} F_{2}^{2} + {\rm tr} F_{3}^{2} - {\rm tr} F_{4}^{2} )^2
\biggr] +
\nonumber
\\
& & -{1\over 24} \Biggl\{ - ({\rm tr} F_{1} + {\rm tr} F_{2} +
{\rm tr} F_{3} + {\rm tr} F_{4} ) \biggl[ {1\over 16} {\rm tr} R^2 
({\rm tr} F_{1} + {\rm tr} F_{2} + {\rm tr} F_{3} + {\rm tr} F_{4}) +
\nonumber
\\
& & \qquad\qquad 
- ({\rm tr} F_{1}^{3} + {\rm tr} F_{2}^{3} + {\rm tr} F_{3}^{3} 
+ {\rm tr} F_{4}^{3} ) \biggr] +
\nonumber
\\
& & \quad\qquad + ({\rm tr} F_{1} + {\rm tr} F_{2} -
{\rm tr} F_{3} - {\rm tr} F_{4} ) \biggl[ {1\over 16} {\rm tr} R^2 
({\rm tr} F_{1} + {\rm tr} F_{2} - {\rm tr} F_{3} - {\rm tr} F_{4}) +
\nonumber
\\
& & \qquad\qquad
- ({\rm tr} F_{1}^{3} + {\rm tr} F_{2}^{3} - {\rm tr} F_{3}^{3} 
- {\rm tr} F_{4}^{3} ) \biggr] +
\nonumber
\\
& & \quad\qquad +3 ({\rm tr} F_{1} - {\rm tr} F_{2} -
{\rm tr} F_{3} + {\rm tr} F_{4} ) \biggl[ {1\over 16} {\rm tr} R^2 
({\rm tr} F_{1} - {\rm tr} F_{2} - {\rm tr} F_{3} + {\rm tr} F_{4})+
\nonumber
\\
& & \qquad\qquad
- ({\rm tr} F_{1}^{3} - {\rm tr} F_{2}^{3} - {\rm tr} F_{3}^{3} 
+ {\rm tr} F_{4}^{3} ) \biggr] +
\nonumber
\\
& & \quad\qquad + 5 ({\rm tr} F_{1} - {\rm tr} F_{2} +
{\rm tr} F_{3} - {\rm tr} F_{4} ) \biggl[ {1\over 16} {\rm tr} R^2 
({\rm tr} F_{1} - {\rm tr} F_{2} + {\rm tr} F_{3} - {\rm tr} F_{4})+
\nonumber
\\
& & \qquad\qquad
- ({\rm tr} F_{1}^{3} - {\rm tr} F_{2}^{3} + {\rm tr} F_{3}^{3} 
- {\rm tr} F_{4}^{3} ) \biggr] \Biggr\} \,.
\nonumber
\ea

In conclusion, we have shown that the {\it crosscap constraint} is
a very powerful tool in orbifold compactifications of the 0B
string theory as well. The freedom in the choice of the
Klein bottle projection allows one to remove both closed and open tachyons
from the spectrum of the open descendants, while maintaining the
consistency of the models. Typically, the ``exotic'' Klein bottle
projection yields only unitary gauge groups, and thus in four dimensions
leads to chiral vacua.

\vskip 24pt
\begin{flushleft} {\large \bf Acknowledgments}
\end{flushleft}

It is a pleasure to thank I. Antoniadis, A. Kumar, G. Pradisi,
K. Ray, S.J. Rey and Ya.S. Stanev.
I am particularly grateful to M. Bianchi and A. Sagnotti. 
This work was supported in part by E.E.C. under TMR
contract ERBFMRX-CT96-0090 and in part by the Italian Ministry of Research.

\end{document}